\documentclass[aps, twocolumn, pra, showpacs, superscriptaddress]{revtex4}
\usepackage{graphicx}
\usepackage{dcolumn}
\usepackage{bm}
\usepackage{amsmath}

\begin{document}

\title{Fidelity susceptibility, scaling, and universality in quantum critical phenomena}

\author{Shi-Jian Gu}\email{sjgu@phy.cuhk.edu.hk}
\affiliation{Department of Physics and ITP, The Chinese University
of Hong Kong, Hong Kong, China}

\author{Ho-Man Kwok}
\affiliation{Department of Physics and ITP, The Chinese University
of Hong Kong, Hong Kong, China}

\author{Wen-Qiang Ning}
\affiliation{Department of Physics and ITP, The Chinese University
of Hong Kong, Hong Kong, China}

\affiliation{Department of Physics, Fudan University, Shanghai
200433, China}

\author{Hai-Qing Lin}
\affiliation{Department of Physics and ITP, The Chinese University
of Hong Kong, Hong Kong, China}

\date{\today }

\begin{abstract}
We study fidelity susceptibility in one-dimensional asymmetric
Hubbard model, and show that the fidelity susceptibility can be used
to identify the universality class of the quantum phase transitions
in this model. The critical exponents are found to be 0 and 2 for
cases of half-filling and away from half-filling respectively.

\end{abstract}





\maketitle

Quantum phase transitions (QPTs) at zero temperature are
characterized by the significant change in the ground state of a
many-body system as a parameter $\lambda$ in the system Hamiltonian
$H(\lambda)$ is varied across a point $\lambda_c$ \cite{Sachdev}.
This primary observation enlightens people to explore the role of
fidelity, a concept emerging from quantum information theory
\cite{Nielsen1}, in the critical phenomena
\cite{HTQuan2006,Zanardi06}. Since fidelity is a measure of
similarity between states, a dramatic change in the structure of the
ground state around a quantum critical point should result in a
great difference between the two ground states on the both sides of
the critical point. Such a fascinating prospect was firstly
confirmed in the 1D XY model where the fidelity shows a narrow
trough at the phase transition point \cite{HTQuan2006,Zanardi06}.
From then on, the fidelity was further used to characterize the QPTs
in fermionic \cite{Pzanardi0606130} and bosonic systems
\cite{Buonsante1}. As fidelity is purely a quantum information
concept, an obvious advantage is that it can be a promising
candidate to characterize the QPT
\cite{PZanardi0701061,PZanardi032109,WLYou07,HQZhou07,LCVenuti07,SChen07}
because no a priori knowledge of the order parameter and the
symmetry of the system is needed. Therefore, these works established
another connection between quantum information theory and condensed
matter physics, in addition to the recent studies on entanglement in
QPTs
\cite{AOsterloh2002,TJOsbornee,GVidal03,SJGuXXZ,SJGUPRL,YChen07}.

The fidelity actually reflects the response of the ground state to a
small change of the driving parameter. Zanardi {\it et al}.
introduced the Riemannian metric tensor \cite{PZanardi0701061}
inherited from the parameter space to denote the leading term in the
fidelity, and argued that the singularity of this metric is in
correspondence with the QPTs. While You {\it et al} introduced
another concept, so-called fidelity susceptibility (FS)
\cite{WLYou07}, and established a general relation between the
leading term in the fidelity and the structure factor of the driving
term in the Hamiltonian. This relation implies that the fidelity may
not have singular behavior in those transitions of infinite order,
such as Kosterlitz-Thouless (KT) transitions \cite{JMKosterlitz73}.

In this work, we study the FS in 1D asymmetric Hubbard model (AHM)
\cite{GFath95}, and show that the FS can be used to characterize the
universality class \cite{HEStanley99} in quantum critical phenomena.
The intrinsic relation between the FS and the Landau's
symmetry-breaking theory (LSBT) is firstly clarified by a simple QPT
occurred in a well-studied 1D transverse-field Ising model. Then we
mainly focus on the critical behavior of the FS in the 1D AHM. Since
the AHM can be used to describe a mixture of two species of
fermionic atoms in an optical lattice, which is able to be realized
by recent experiments on the cold atoms \cite{CChin04}, the model
itself is of current research interest
\cite{CAMacedo02,DUeltschi04,VJEmery87,MACazalilla05,SJGu05,ZGWang07}.
We find that the critical exponents of the FS take the value of 0
and 2 for cases of half-filling ($n=1$) and away from half-filling
($n=2/3$) respectively.

To begin with, we consider a general Hamiltonian of quantum
many-body systems, i.e.
\begin{eqnarray}
H(\lambda, h)=H_0 +\lambda H_I + h M, \label{eq:Hamitonian}
\end{eqnarray}
where $H_I$ is the driving Hamiltonian with the strength $\lambda$,
and $M$ is a potential order parameter and $h$ is the corresponding
external field. Without loss of generality, we first set $h=0$.
Following Ref. \cite{Zanardi06}, the fidelity is defined as the
overlap between two ground states $|\Psi_0(\lambda)\rangle$ and
$|\Psi_0(\lambda+\delta\lambda)\rangle$, that is
\begin{eqnarray} F (\lambda,
\delta\lambda)=|\langle\Psi_0(\lambda)|\Psi_0(\lambda+\delta\lambda)\rangle|.
\label{eq:fidedifintion}
\end{eqnarray}
Then the FS is just the most relevant term in the fidelity, and
mathematically is related to the structure factor of the driving
term $H_I$ \cite{WLYou07}, which denotes the fluctuation caused by
the driving parameter. For example, if we extend the fidelity to the
thermal state \cite{PZanardi032109}, the FS is just the specific
heat or the magnetic susceptibility \cite{PZanardi0701061,WLYou07}
if we choose the driving parameter as temperature or magnetic field
respectively.

Compared with ordinary phase transitions, we can also extract two
exponents (denoted as $\alpha, \gamma$) from the FS in quantum
critical phenomena if we choose the driving parameter as $\lambda$
and $h$ (if the order parameter is known) respectively (Here, the
only condition is that the QPT should belong to the type of Landau's
transition, otherwise $\alpha=0$ and $\gamma$ is not well defined).
Then the fidelity susceptibilities driven by two terms in the
Hamiltonian (\ref{eq:Hamitonian}) scale like
\begin{eqnarray}
\frac{\chi_{F(\lambda)}(\lambda)}{N}\propto\frac{1}{|\lambda_c-\lambda|^\alpha},\;\;\;
\frac{\chi_{F(h=0)}(\lambda)}{N}\propto\frac{1}{|\lambda_c-\lambda|^\gamma},
\label{eq:criticalexpdef}
\end{eqnarray}
respectively, around the critical point $\lambda_c$ in the
thermodynamic limit. As a simple application, we take the
well-studied model, i.e. 1D transverse-field Ising model, as an
example,
\begin{eqnarray}
H_{\rm Ising}=\sum_{j}\left[\sigma_j^z\sigma_{j+1}^z +\lambda
\sigma^x_j + h\sigma_j^z\right], \label{eq:Hamiltonian_Ising}
\end{eqnarray}
where $\sigma$ is Pauli matrix. In Ref. \cite{Zanardi06}, it was
obtained $\alpha=1$. On the other hand, if we consider $h$ as the
driving parameter, we find $\gamma=7/4$, which is the same as the
$\gamma$ of the 2D Ising model \cite{PSen00}. Take into account the
exponent $\beta=1/8$ of the order parameter $\sigma^z$, we then have
$\alpha + 2\beta +\gamma=3$, which is slightly different from the
usual $\alpha + 2\beta +\gamma=2$ in 2D Ising model. We interpret
this difference as one more differentiation is made in the FS of the
ground state than the specific heat at finite temperatures. That is,
the FS is related to the second order derivative of the ground state
energy with respect to the driving parameter \cite{WLYou07}, while
for the specific heat, it is simply $dE(T)/dT$ where $E(T)$ is the
internal energy. Therefore, the phase transition here still belongs
to the same universality class of the 2D Ising model. As a brief
conclusion, the relation between the FS and the LSBT is
straightforward. Once the driving term and order parameter are
given, the universality classes are simply described by the critical
exponents of FS.


The LSBT is established on the order parameter, whose non-vanishing
behavior results from the broken symmetry and long-range order. For
KT transitions, both broken symmetry and long-range order are
absent, hence no local order parameter is concerned. Consider again
the original definition of the FS, i.e.
\begin{eqnarray}
\chi_F(\lambda)=\sum_{n\neq
0}\frac{|\langle\Psi_n(\lambda)|H_I|\Psi_0(\lambda)
\rangle|^2}{[E_n(\lambda)-E_0(\lambda)]^2} \label{eq:Fidelityexp2}
\end{eqnarray}
where $|\Psi_n(\lambda)\rangle$ satisfies
$H(\lambda)|\Psi_n(\lambda)\rangle =E_n |\Psi_n(\lambda)\rangle$ and
defines a set of orthogonal complete basis in the Hilbert space. For
the KT transition, despite of the vanishing energy gap, there is
still no singularity in the FS as matrix elements
$\langle\Psi_n(\lambda)|H_I|\Psi_0(\lambda) \rangle$ also vanish at
the same time. However, the appearance of the power-law decay
behavior describes the stronger fluctuation around the critical
point. This point directly leads to that the FS, which also denotes
the fluctuation of the driving term, might reach a maximum near the
critical point, though the maximum point might not be the critical
point, as has been observed in the 1D Hubbard model\cite{WLYou07}.

\begin{figure}
\includegraphics[width=8cm]{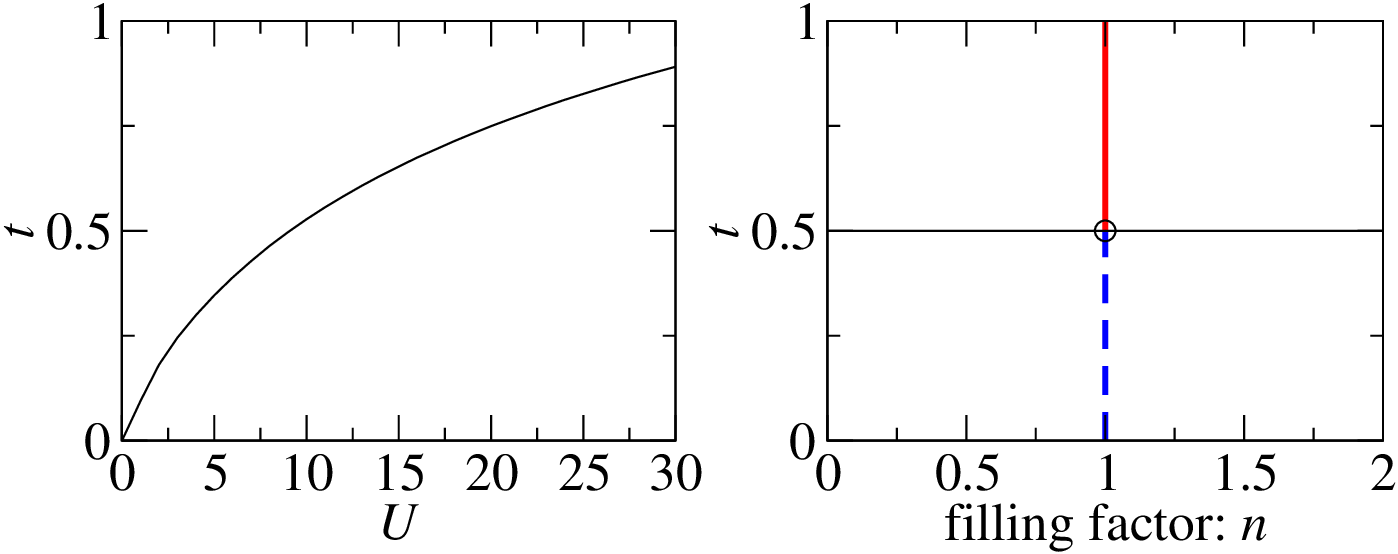}
\caption{ (color online) The schematic phase diagram of the AHM.
LEFT: the phase diagram defined on the $U-t$ plane which shows a KT
transition at half-filling and Landau's transition away from
half-filling. RIGHT: the phase diagram defined on the $n-t$ plane,
in which the transition along the middle line is of KT type and the
thin solid lines at both sides are of Landau's type. The phase
diagram has a mirror symmetry about the line $n=1$ due to the
particle hole symmetry in the model.} \label{figure_sphase}
\end{figure}

To confirm this physical intuition, we now focus on the 1D AHM,
whose Hamiltonian reads
\begin{eqnarray} H_{\rm AHM}=-\sum_{j=1}^{L}\sum_{\delta=\pm 1}
\sum_\sigma t_\sigma c^\dagger_{j,\sigma}c_{j+\delta, \sigma}+U
\sum_{j=1}^L n_{j, \uparrow}n_{j, \downarrow},
\label{eq:Hamiltonian_AHM}
\end{eqnarray}
where $c^\dagger_{j,\sigma}$ and
$c_{j,\sigma},\sigma=\uparrow,\downarrow$ are creation and
annihilation operators for electrons with spin $\sigma$ at site $j$
respectively, $n_\sigma=c_\sigma^\dagger c_\sigma$, $t_\sigma$ is
$\sigma$-dependent hoping integral, and $U$ denotes the strength of
on-site interaction. In this model, the Hamiltonian has
U(1)$\otimes$U(1) symmetry for general $t_\sigma$, and the atoms
number $N_\uparrow=\sum_j n_{j,\uparrow}, N_\downarrow=\sum_j n_{j,
\downarrow}$ are conserved respectively. The total number of atoms
is given by $N=N_\uparrow+N_\downarrow$, and the filling factor is
$n=N/L$. For simplicity, we reset $t=t_\downarrow/t_\uparrow$, and
$U$ to be $U/t_\uparrow$.

The schematic phase diagram of the AHM is shown in Fig.
\ref{figure_sphase}, which can be understood from its two limiting
cases, the Hubbard model \cite{Hubbard} ($t_\uparrow=t_\downarrow$)
and the Falicov-Kimball (FK) model \cite{LMFalicov69,TKennedy86}
($t_\downarrow=0$). At half-filling, both the Hubbard model and the
FK model are in a spin-density-wave state. The difference is that in
the Hubbard region, the system renormalizes to the Heisenberg fixed
point, while in the FK region, it belongs to the Ising fixed point.
The QPT occurred between these two classes belongs to the KT type in
the 1D system because the correlation function at both side is of
power-law decay \cite{GFath95} and no local order parameter is well
defined. While away from half filling, the system becomes an ideal
conductor and is in the state of density wave in the Hubbard region,
but it is in a phase separation in the FK region. In the phase
separation region, the translational symmetry is broken, and the
down-spin electrons congregate together; then $\langle n_\downarrow
\rangle$ plays a role of the order parameter. So the phase
transition is of Landau's type \cite{SJGu05,ZGWang07}.

\begin{figure}
\includegraphics[width=8cm]{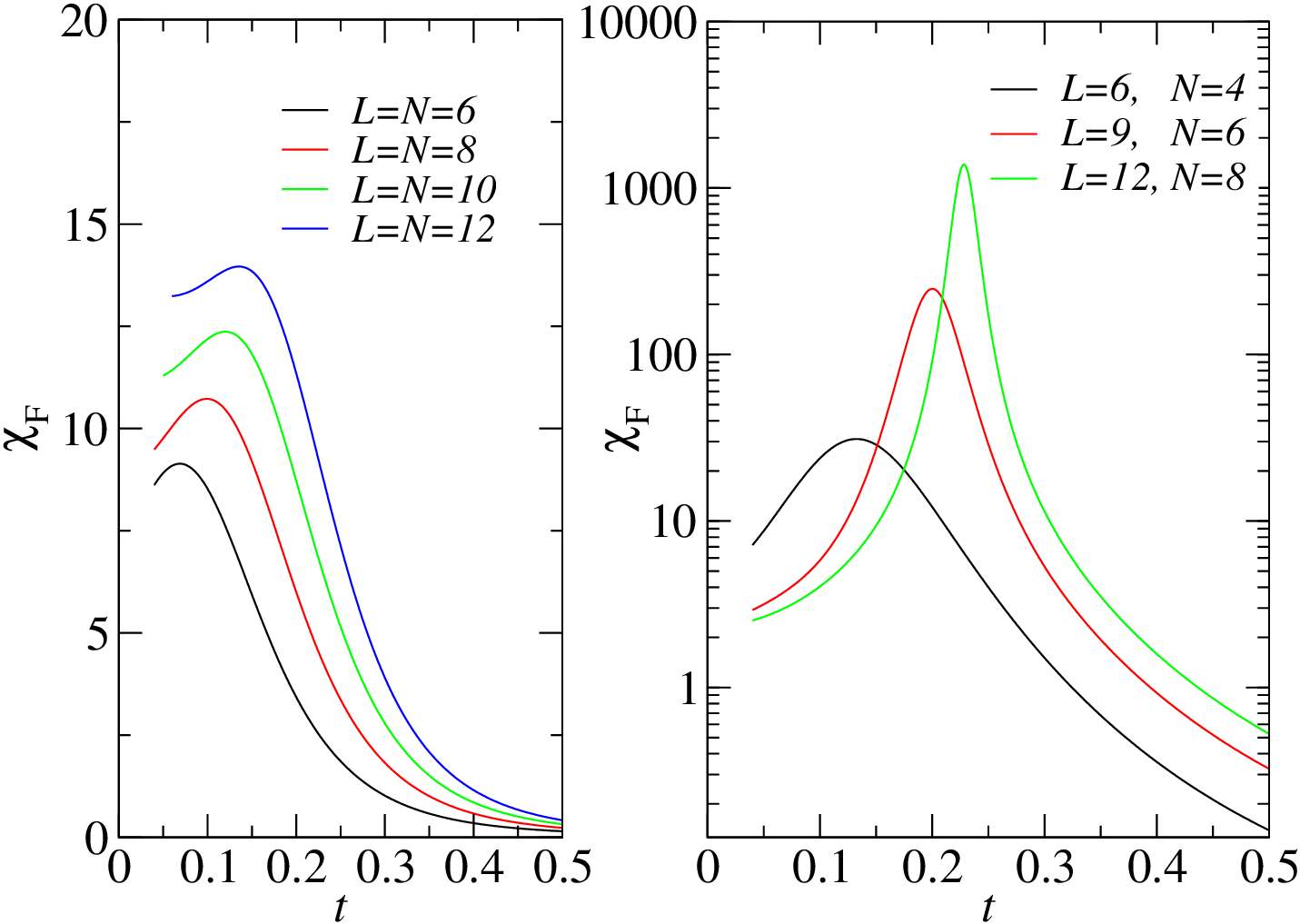}
\caption{ (color online) The scaling behavior of the FS as a
function of $t$ for the cases of $n=1$ (LEFT) and $n=2/3$ (RIGHT).
Here $U=10$.\\} \label{figure_fdsu10}
\end{figure}

In order to quantify the change of the ground state during the
evolution of $t$, we define the fidelity as $F(t, \delta t)=|\langle
\Psi(t)|\Psi(t+\delta t)\rangle|$. The corresponding FS is
$\chi_{F(t)}(t)=-2\lim_{\delta t\rightarrow 0}\ln F(t, \delta
t)/\delta t^2$. To avoid the ground state level crossing, we choose
the periodic or antiperiodic boundary conditions for systems with
$4l+2$ or $4l$ electrons respectively. We first look at two special
cases, i.e. the FS at half-filling ($n=1$) and away from
half-filling ($n=2/3$), and both with a given interaction $U=10$.
The numerical results of different system sizes are presented in
Fig. \ref{figure_fdsu10}. For both cases, the FS reaches a maximum
point at a certain position $t_{\rm max}$. The difference is that
for $n=2/3$ case, $\chi_{F(t)}(t=t_{\rm max})$ diverges with
increasing system size; while for $n=1$, $\chi_{F(t)}(t=t_{\rm max})
\propto L$. The former behavior clearly denotes a Landau's
transition, and the latter is KT transition.

\begin{figure}
\includegraphics[width=8cm]{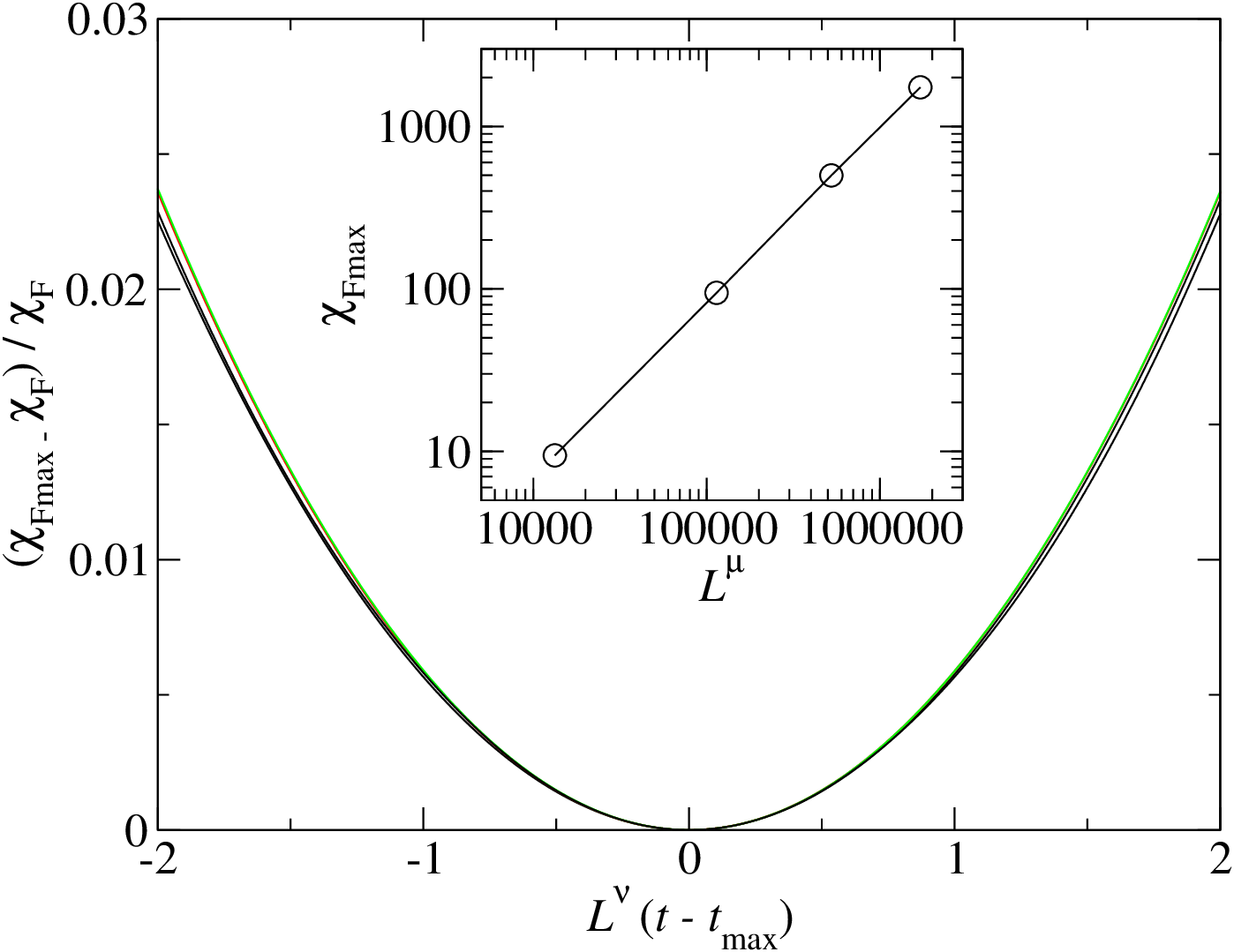}
\caption{ (color online) The finite size scaling is performed for
the case of power-law divergence for the case of $U=30$, $n=2/3$,
and system sizes $L=6, 9, 12, 15$. The FS, consider as a function of
the system size and the driving parameter is a function of
$L^\nu(t-t_{\rm max})$ only. Here the critical exponent is
$\nu\simeq 2.65$. The inset denotes the scaling behavior of
$\chi_{F}(t=t_{\rm max})$. The straight line is of slope 1 in
logarithmic scale, and $\mu\simeq 5.3$.} \label{figure_scale_U10}
\end{figure}

To study the critical behavior around the critical point, finite
scaling analysis is needed. According to the definition of the
critical exponents in Eq. (\ref{eq:criticalexpdef}), we introduce
the following scaling behavior for the FS
\begin{eqnarray}
\chi_{F(t)}(t, L) = \frac{A}{L^{-\mu} + B(t-t_{\rm max})^\alpha},
\label{eq:scalingfunction}
\end{eqnarray}
where $A, B$ are constants independence of $L$ and $t$. Such a
finite size scaling leads to that the rescaled  FS
$[\chi_{F(t)}(t=t_{\rm max})-\chi_{F(t)}(t)]/\chi_{F(t)}(t)$ is a
simple function of the rescaled driving parameter $L^\nu (t-t_{\rm
max})$. This function is universal and does not depend on system
sizes, as shown in Fig. \ref{figure_scale_U10} for the case of
$U=30$, in which numerical results obtained from various system
sizes fall onto a single line. On the other hand, if $t=t_{\rm max}$
which approaches to $t_c$ like $t_{\rm max}-t_c\propto L^{-2}$, the
maximum value of the FS diverges with increasing system size as:
$\chi_{F(t)}(t=t_{\rm max})\propto L^\mu$. Then the exponent $\nu$
together with $\mu$ determines the critical exponent $\alpha$ in Eq.
(\ref{eq:criticalexpdef}). For the present case, we find $\nu\simeq
2.65$ and $\mu\simeq5.3$, hence $\alpha=\mu/\nu=2$. That is, around
the critical point, the FS for the case of $n=2/3$ scales like
\begin{eqnarray}
\frac{\chi_{F(t)}(t)}{L}\propto \frac{1}{|t-t_c|^2},
\end{eqnarray}
which clearly differs from the Ising model. For the Ising model [Eq.
(\ref{eq:Hamiltonian_Ising})], only $Z_2$ symmetry is broken when
the phase transition occurs; while in the AHM, the translational
symmetry is broken in the phase separation region. So they belong to
different universality classes.

\begin{figure}
\includegraphics[width=8cm]{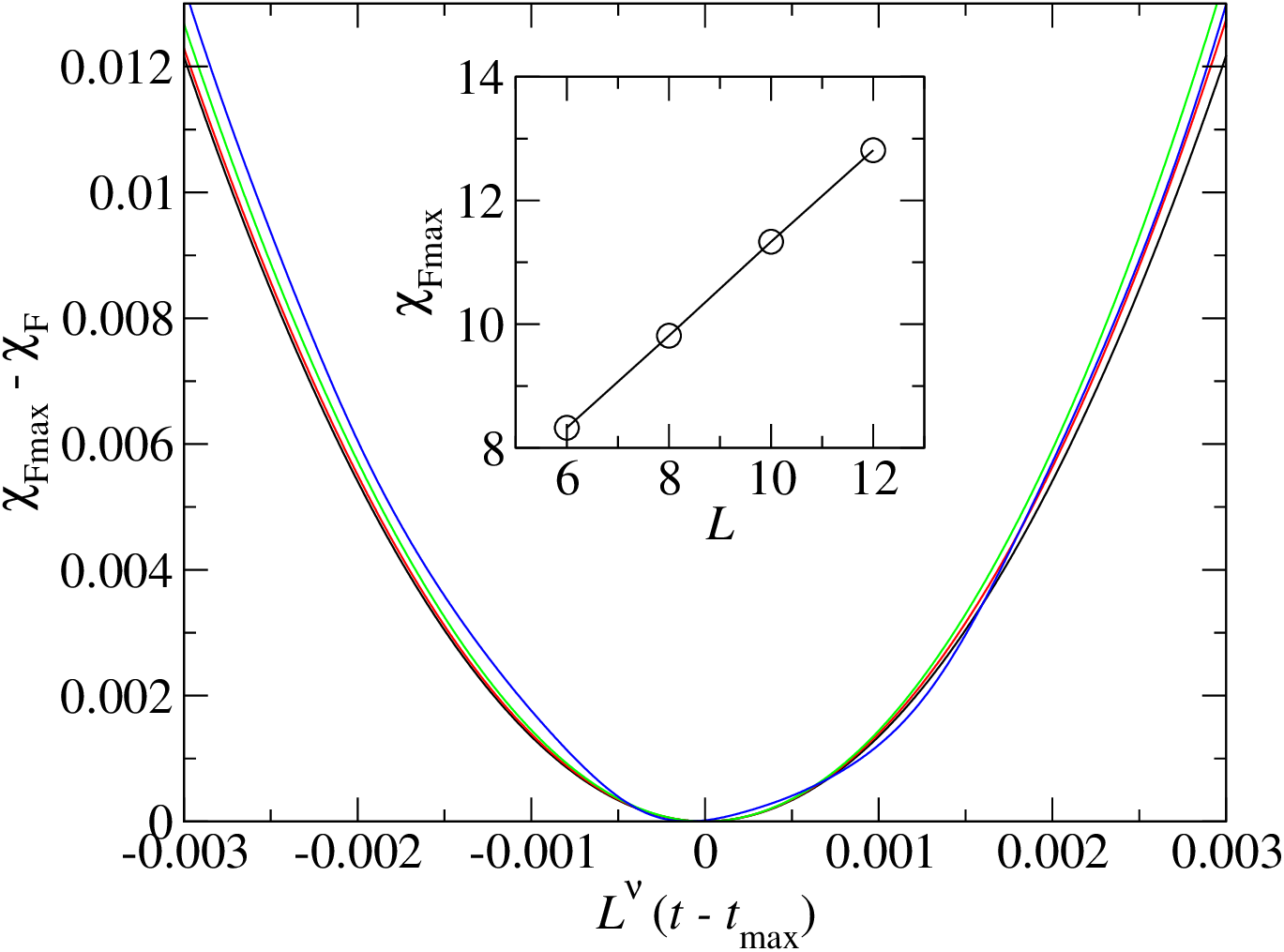}
\caption{ (color online) The similar finite size scaling is
performed for the case of KT transition occurred for various system
size $L=6, 8, 10, 12$ at $n=1$ and $U=30$. The FS, consider as a
function of the system size and the driving parameter is a function
of $L^\nu(t-t_{\rm max})$ only. Here the critical exponent is
$\nu\simeq -0.25$. The inset denotes the scaling behavior of
$\chi_{F}(t=t_{\rm max})$. } \label{figure_scale_U10_1o1}
\end{figure}

The KT transition occurs at half-filling $n=1$. The corresponding
finite scaling analysis for the case $U=30$ is presented in Fig.
\ref{figure_scale_U10_1o1}. The maximum point of the FS is
proportional to the system length. This is consistent with our
aforementioned understanding. On the other hand, the rescaled FSs
for various system sizes fall onto a single line, which is a
function of $L^\nu(t-t_{\rm max})$ with critical length exponent
$\nu=-1/4$. We find that the FS around the maximum point like:
\begin{eqnarray}
\chi_{F(t)}(t)\simeq 3.855 + 0.7478 L + 1349.9 L^{-1/2} (t-t_{\rm
max})^2,
\end{eqnarray}
around $t_{\rm max}$. As expected, there is no singularity in
$\chi_{F(t)}(t)$. Clearly, the maximum behavior becomes weak with
the increasing system size, as we can also find in Fig.
\ref{figure_fdsu10}. Then the FS becomes flat in the FK region, the
relative larger FS is due to the power-law behavior the correlation
function. On the other hand, motivated by the KT transition occurred
in quantum XY model \cite{HQDing92}, we infer that the very steep
decreasing point of the FS in Fig. \ref{figure_fdsu10} is more close
to the critical point. Therefore, we perform $1/L$ finite scaling
analysis for the minimum point of $d\chi_{F(t)}(t)/dt$, and find
$t_c \simeq 0.308, 0.313, 0.317$ for $U=10, 20, 30$ respectively.
The results are very close to those obtained by density matrix
renormalization group method \cite{GFath95}.

A similar analysis can be carried on for other filling conditions.
The power-law divergence of the FS always exists in other filling
conditions except when $n=1$. On the other hand, due to the
particle-hole symmetry in the AHM, the FS takes the same value for
$n$ and $2-n$ ($n<2$), which satisfies the same scaling behavior.
Therefore, the phase diagram in the left picture of Fig.
\ref{figure_sphase} has a mirror symmetry about the line $n=1$. Take
into account the fact that even a single hole doping might lead to
the instability of the density wave state in the infinite $U$ limit
\cite{SJGu05}, the KT transition only happens at the half-filling
condition. So the circle point in Fig. \ref{figure_sphase} is
expected to be a quar-critical point in the phase diagram, and the
FS just signals the transition type along the critical lines.

In conclusion, we have shown that the FS, as the leading term in the
fidelity between two ground states separated by a slightly
difference in parameter space, can be used to characterize the
universality class in quantum critical phenomena. Since the FS is
related to the structure factor of the driving term in the
Hamiltonian, the relation between the LSBT and FS is linked up. Its
critical exponent then is naturally suitable for the classification
of universality. We elucidate this relation by the simple QPT
occurred in the 1D transverse-field Ising model. Furthermore,
despite no singularity appearing in the FS when crossing a KT
transition point, the stronger fluctuation might makes the FS reach
a maximum near to the critical point. We then studied the FS in the
1D AHM, and shown that the FS can help us to identify both types of
phase transition in this model. The critical exponent $\alpha$ for
the Landau's type transition is calculated with finite size
analysis, and is found to be $\alpha=2$ for $n=2/3$ case. While for
the KT transition, $\alpha=0$.

\emph{Note added.} Recently, the work on the scaling behavior of the
FS in other model appeared \cite{LCVenuti07}.

This work is supported by RGC Grant CUHK 400906.


\begin{thebibliography}{99}

\bibitem{Sachdev} S. Sachdev, {\it Quantum Phase Transitions} (Cambridge University Press,
Cambridge, England, 1999).

\bibitem{Nielsen1}
M. A. Nilesen and I. L. Chuang, {\it Quantum Computation and Quantum
Information} (Cambridge University Press, Cambridge, England, 2000)


\bibitem{HTQuan2006}
H. T. Quan, Z. Song, X. F. Liu, P. Zanardi, and C. P. Sun, Phys.
Rev. Lett. {\bf 96}, 140604 (2006).

\bibitem{Zanardi06} P. Zanardi and N. Paunkovi¡äc, Phys. Rev. E \textbf{74}, 031123
(2006).

\bibitem{Pzanardi0606130}
P. Zanardi, M. Cozzini, and P. Giorda, arXiv: quant-ph/0606130; M.
Cozzini, P. Giorda, and P. Zanardi, arXiv: quant-ph/0608059; M.
Cozzini, R. Ionicioiu, and P. Zanardi, arXiv: cond-mat/0611727.

\bibitem{Buonsante1} P. Buonsante1 and A. Vezzani, Phys. Rev. Lett. \textbf{98}, 110601
(2007).

\bibitem{PZanardi032109}
P. Zanardi, H. T. Quan, X. Wang, and C. P. Sun, Phys. Rev. A {\bf
75}, 032109 (2007).

\bibitem{PZanardi0701061}
P. Zanardi, P. Giorda, and M. Cozzini, arXiv: quant-ph/0701061v1.

\bibitem{WLYou07}
W. L. You, Y. W. Li, S. J. Gu,  arXiv:quant-ph/0701077.

\bibitem{HQZhou07}
H. Q. Zhou, J. P. Barjaktarevic, arXiv: cond-mat/0701608; H. Q.
Zhou, J. H. Zhao, B. Li, arXiv:0704.2940; H. Q. Zhou,
arXiv:0704.2945.


\bibitem{LCVenuti07}
L. C. Venuti, P. Zanardi, arXiv:0705.2211.

\bibitem{SChen07}
S. Chen, L. Wang, S. J. Gu, Y. Wang, arXiv:0706.0072.


\bibitem{AOsterloh2002}
A. Osterloh, Luigi Amico, G. Falci and Rosario Fazio, Nature {\bf
416}, 608 (2002).

\bibitem{TJOsbornee}
T. J. Osborne and M.A. Nielsen, Phys. Rev. A {\bf 66}, 032110(2002).

\bibitem{GVidal03}
G. Vidal, J. I. Latorre, E. Rico, and A. Kitaev, Phys. Rev. Lett.
90, 227902 (2003).

\bibitem{SJGuXXZ}
S. J. Gu, H. Q. Lin, and Y. Q. Li, Phys. Rev. A {\bf 68}, 042330
(2003).

\bibitem{SJGUPRL}
S. J. Gu, S. S. Deng, Y. Q. Li, H. Q. Lin, Phys. Rev. Lett. {\bf
93}, 086402 (2004).

\bibitem{YChen07}
Y. Chen, Z. D. Wang, Y. Q. Li, and F. C. Zhang, Phys. Rev. B
\textbf{75}, 195113 (2007).

\bibitem{JMKosterlitz73}
J. M. Kosterlitz and D. J. Thouless, J. Phys. C {\bf 6}, 1181(1973).

\bibitem{HEStanley99}
H. E. Stanley, Rev. Mod. Phys. \textbf{71}, S358 (1999).

\bibitem{GFath95}
G. F\'{a}th, Z. Doma\'{n}ski, and R. Lema\'{n}ski, Phys. Rev. B {\bf
52}, 13910 (1995); Z. Doma\'{n}ski, and R. Lema\'{n}ski, G.
F\'{a}th, J. Phys. Condens. Matter {\bf 8}, L261 (1996).

\bibitem{CChin04} 
For examples: C. Chin  {\it et al}, Science {\bf 305}, 1128 (2004); %
J. Kinast, {\it et al}, Science {\bf 307}, 1296 (2005); %
M. K\"{o}hl, {\it et al}, Phys. Rev. Lett. {\bf 94}, 080403 (2005); %
M. Bartenstein {\it et al}, Phys. Rev. Lett. {\bf 94}, 103201
(2005).


\bibitem{CAMacedo02}
C. A. Macedo and A. M. C. de Souza, Phys. Rev. B {\bf 65}, 153109
(2002).

\bibitem{DUeltschi04}
D. Ueltschi, J. Stat. Phys. {\bf 116}, 681 (2004).

\bibitem{VJEmery87}
V. J. Emery, Phys. Rev. Lett. {\bf 58}, 2794 (1987).


\bibitem{MACazalilla05}
M. A. Cazalilla, A. F. Ho, and T. Giamarchi, Phys. Rev. Lett. {\bf
95}, 226402 (2005).

\bibitem{SJGu05}
S. J. Gu, R. Fan, and H. Q. Lin, arXiv: cond-mat/0601496.

\bibitem{ZGWang07}
Z. G. Wang, Y. G. Chen, and S. J. Gu, Phys. Rev. B {\bf 75}, 165111
(2007).

\bibitem{PSen00}
P. Sen, Phys. Rev. E \textbf{63}, 016112 (2000).

\bibitem{Hubbard}
J. Hubbard, Proc. R. Soc. London A {\bf 276}, 238 (1963); %
E. H. Lieb and F. Y. Wu, Phys. Rev. Lett. {\bf 20}, 1445 (1968).

\bibitem{LMFalicov69}
L. M. Falicov and J. C. Kimball, Phys. Rev. Lett. {\bf 19}, 997
(1969).

\bibitem{TKennedy86}
T. Kennedy and E. Lieb, Physica A 138:320 (1986).

\bibitem{HQDing92}
H. Q. Ding, Phys. Rev. B {\bf 45}, 230 (1992).

\end{thebibliography}
\end{document}